\documentclass[
    ,final            
  ]
  {aipproc}

\layoutstyle{6x9}

\begin{document}

\title{Hadron Spectroscopy - A 2005 Snapshot}


\author{Jonathan L. Rosner}{
  address={Enrico Fermi Institute and Department of Physics,
  University of Chicago, 5640 S. Ellis Avenue, Chicago IL 60637 USA}
}

\begin{abstract}
  Some aspects of hadron spectroscopy are reviewed as of summer 2005.
\end{abstract}

\maketitle


\section{INTRODUCTION}

Hadron spectroscopy plays a valuable role in particle physics.  It
was crucial in validating quantum chromodynamics (QCD) and the quark
substructure of matter.  It provides a stage for understanding nonperturbative
techniques, not only in QCD but potentially elsewhere in physics.
Hadron spectra are crucial in separating electroweak physics from
strong-interaction effects, as in charm and beauty decays.  Quarks and leptons
have an intricate level and weak coupling structure for which
we have no fundamental understanding.  Sharpening spectroscopic techniques may
help solve this problem.

I review hadron spectroscopy with emphasis on new results from
heavy-quark studies, discussing no-quark states (glueballs), light-quark
states, charmed hadrons, charmonium, beauty hadrons, and the $\Upsilon$
family, concluding with a homework assignment.

\section{GLUEBALLS AND PARTICLES LOOKING LIKE THEM}

In QCD, quarkless states (``glueballs'') may be constructed from pure-glue
configurations.  If $F^a_{\mu \nu}$ is the gluon field-strength tensor,
one can form $J^{PC} = 0^{++}$ states as $F^a_{\mu \nu} F^{a \mu \nu}$,
$0^{-+}$ as $F^a_{\mu \nu} \tilde{F}^{a \mu \nu}$, and higher-spin
configurations using derivatives or more than two gluon fields.  All such
states should be flavor-singlet with isospin $I=0$.  Although equal
coupling to light nonstrange and strange quarks has usually been assumed,
$s \bar s$ couplings could be favored \cite{Chanowitz:2005du}.  Lattice QCD
calculations predict the lowest glueball to be $0^{++}$ with $M \simeq
1.7$ GeV.  Many other $I=0$ levels, e.g., $q \bar q$, $q \bar q g$
($g$ = gluon), $q q \bar q \bar q, \ldots$, can mix with such a state.  One
must study $I=0$ levels and their mesonic couplings to separate out glueball,
$n \bar n \equiv (u \bar u + d \bar d)/ \sqrt{2}$,
and $s \bar s$ components.  Understanding the rest of the {\it flavored}
$q \bar q$ spectrum for the same $J^P$ thus is crucial.

\begin{figure}
\mbox{\includegraphics[width=0.4\textwidth]{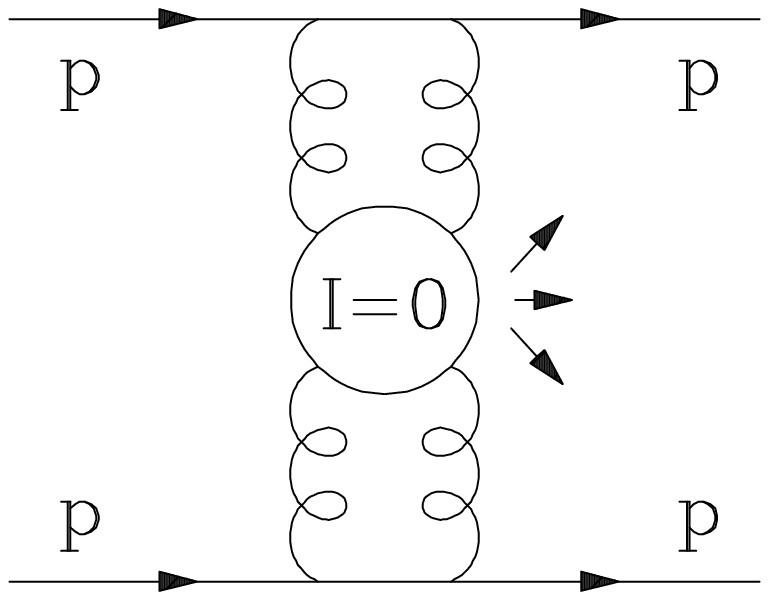} \hskip 0.3in
      \includegraphics[width=0.45\textwidth]{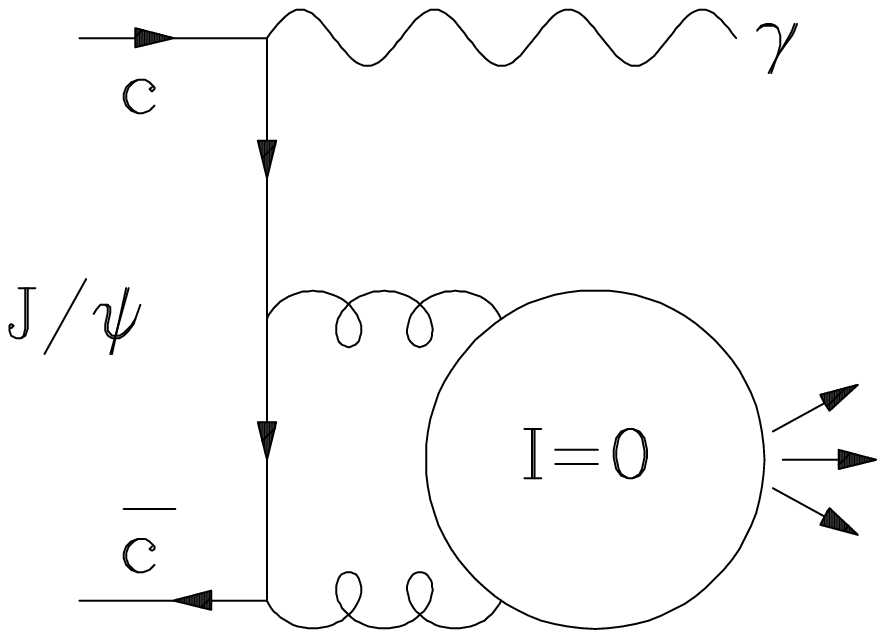}}
\caption{Two mechanisms for producing glueballs.  Left:  Central production in
hadron-hadron double Pomeron exchange; Right: Radiative quarkonium decay.
\label{fig:prod}}
\end{figure}

Two mechanisms for producing glueballs are shown in Fig.\ \ref{fig:prod}.  In
double pomeron exchange (as utilized, for example, in CERN Experiment WA-102
\cite{Barberis:2001bs}), the centrally produced state is expected to have $I =
0$. Radiative $J/\psi$ decay is about 10\% of three-gluon decay and also should
populate $I=0$ states if the photon couples to the charmed quark.  The
BES Collaboration has obtained a sample of 58 million $J/\psi$ decays, and
the CLEO-c detector hopes to record an even larger number.  Other processes
for glueball production include $\pi N \to \pi \pi N$ and $\bar p p$
annihilation \cite{Amsler:2003bq}.
One can diagnose the flavor of neutral states $X$ by noting that $J/\psi \to
(\omega,\phi) X$ favors (nonstrange,strange) $X$, with $X \to \gamma+(\rho,
\omega,\phi)$ testing the flavor content of $X$.  A glueball is expected to
have a small decay width to $\gamma \gamma$.

One pattern of glueball masses is predicted in Ref.\ \cite{Campbell:1997}.
The lowest state has $J^{PC} = 0^{++}$ and mass around 1.7 GeV.  The
next-lowest are $2^{++}~(M \simeq 2.4$ GeV) and $0^{-+}~(M \simeq 2.6$ GeV).
These predictions are difficult both to make and to test, since mixing with
decay channels and $(q \bar q)^n~(n \ge 1)$ configurations can occur.  The best
prospects are to sort out the $0^{++}$ $I=0$ mesons \cite{Eidelman:2004wy},
for which several candidates exist.

The $f_0(600)$ (with $\Gamma \ge 500$ MeV), a possible QCD analogue of the
Higgs boson, appears to be a dynamically induced $\pi \pi$ resonance.  It can
be generated from $\pi \pi$ scattering using just current algebra, crossing
symmetry, and unitarity \cite{Brown:1971ea}.  It also can be interpreted as
a candidate for a light $qq \bar q \bar q$ state \cite{Jaffe:2003sg}.
The $f_0(980)$ is a narrow resonance with $\Gamma = 40$--100 MeV) which appears
to be strongly correlated with $K \bar K$ threshold.  It has been interpreted
variously as $s \bar s$, $s \bar s n \bar n$, an $I=0~K \bar K$ molecule, or a
mixture of these states.  For a recent discussion, see \cite{Achasov:2005ub}.

Three states have been proposed as mixtures of a glueball and
$^3P_0$ $q \bar q$ states \cite{Close:2005vf}.  (1) The $f_0(1370)$, with
$\Gamma = 200$--500 MeV, viewed mainly as $n \bar n$,
is seen in $J/\psi \to \gamma \pi^+ \pi^-$. (2) The $f_0(1500)$,
with $\Gamma \simeq 100$ MeV, is not seen in $J/\psi \to \gamma \pi^+ \pi^-$
but appears in $\bar p p$ annihilations and central $K \bar K$ production.
Ref.\ \cite{Close:2005vf} consider it to be mainly a glueball $G$.  (3) The
$f_0(1700)$, with $\Gamma = 250\pm100$ MeV, is considered in Ref.\
\cite{Close:2005vf} to be mainly $s \bar s$ since it is prominent in $J/\psi
\to \gamma K \bar K$.  It could be a glueball if $ \to s \bar s$ couplings are
enhanced \cite{Chanowitz:2005du}.  The scheme of Ref.\ \cite{Close:2005vf} is
consistent with production in $pp$, $J/\psi \to \gamma f_0$, and $J/\psi \to V
f_0$ [$V=(\rho,\omega,\phi)$] processes but needs to be tested via
$f_0 \to V \gamma$.

The $2^{++}$ candidates listed in \cite{Eidelman:2004wy} are $f_2(1275)$,
$f'_2(1525)$, $f_2(1950)$, $f_2(2010)$, $f_2(2300)$, and $f_2(2340)$.  The
first two are well-established $1^3P_2(q \bar q)$ states.  Higher states
with masses within 100 MeV of one another may not be distinct if their
observed properties depend on production.  The $f_2(1950-2010)$ are in the
range expected for radial $q \bar q$ excitations.  The highest-lying ones are
in the expected glueball range, however.  Their flavor diagnosis via decay
modes or suppressed $\gamma \gamma$ production is crucial.

$0^{-+}$ candidates \cite{Eidelman:2004wy} are $\eta(548)$, $\eta'(958)$,
$\eta(1295)$, $\eta(1405)$, and $\eta(1475)$, while a state $X(1835)$, decaying
to $\eta' \pi \pi$, has recently been reported in $J/\psi \to \gamma X$
\cite{Ablikim:2005um}.  There is room for glueball content of these states
but two of the three between $\sim 1.3$ and 1.5 GeV are likely radial
$2^1S_0(q \bar q)$ excitations, while the glue content of $\eta'$ is limited
by the observed $\phi \to \eta' \gamma$ rate \cite{Passeri:2003jc}.
The state $X(1835)$ is consistent
with the enhancement at $\bar p p$ threshold reported earlier by BES in
$J/\psi \to \gamma \bar p p$ \cite{Bai:2003sw} if $J^{PC}(X) = 0^{-+}$.  (The
$\bar p p$ enhancement was also consistent with $0^{++}$ but such a state
cannot decay to $\eta' \pi \pi$.)  It could be a glueball with a large $\bar p
p$ content in its wave function (but is below the lowest expected
$0^{-+}$ mass \cite{Campbell:1997} of 2.6 GeV), or a $\bar p p$ ``baryonium''
resonance \cite{Zhu:2005ns}.

\begin{figure}
\mbox{\includegraphics[width=0.32\textwidth]{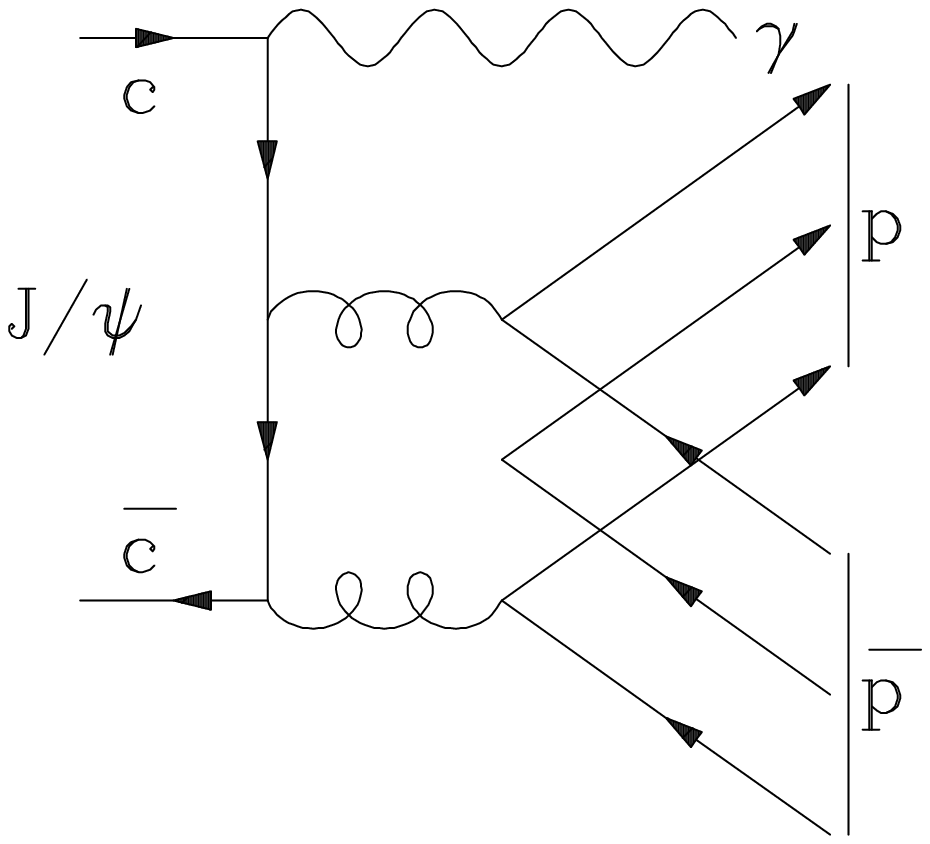} \hskip 0.1in
      \includegraphics[width=0.66\textwidth]{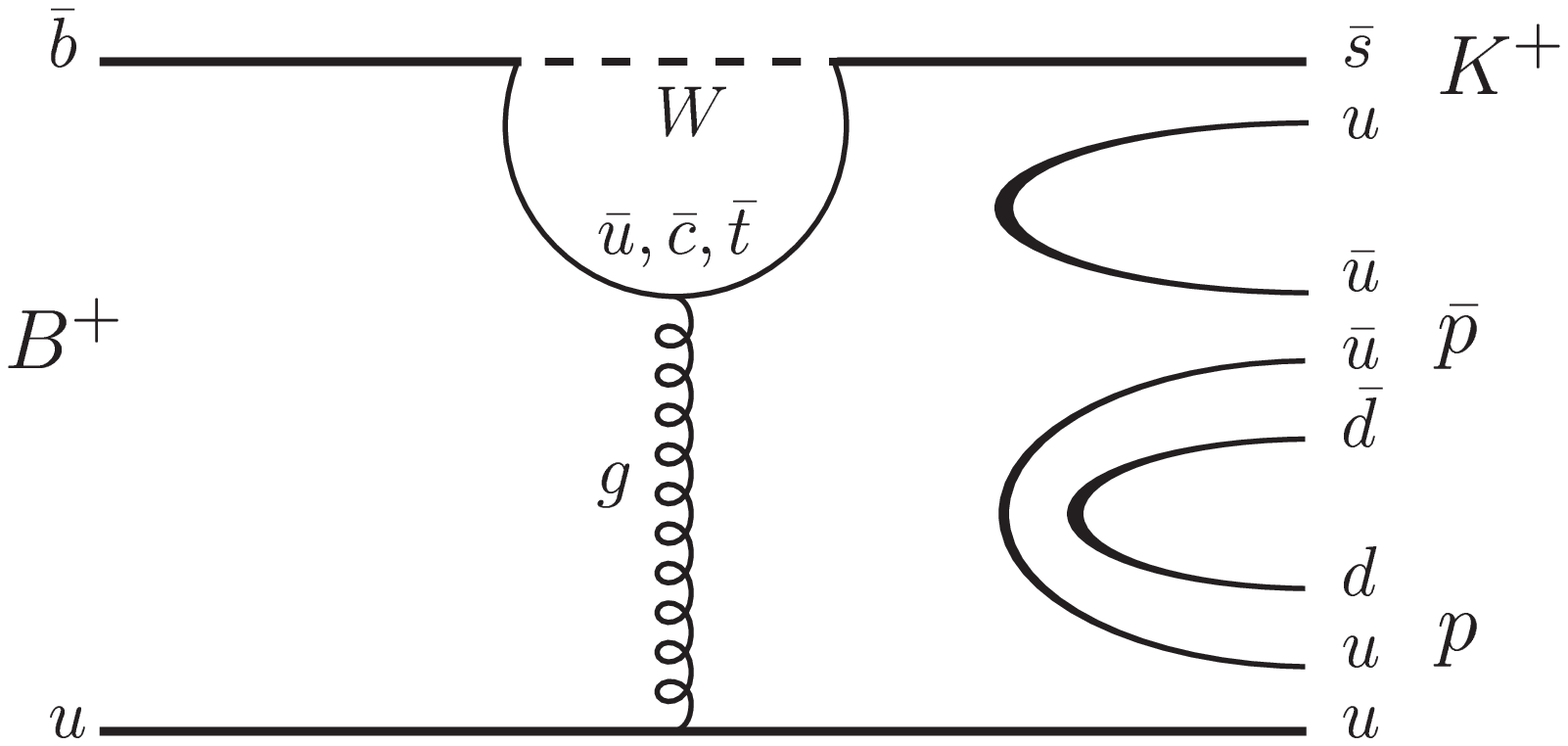}}
\caption{Mechanisms for producing threshold baryon-antibaryon enhancements.
Left:  radiative $J/\psi$ decay; Right: penguin-dominated $B$ decay.
\label{fig:bbbar}}
\end{figure}

Baryon-antibaryon enhancements near threshold appear not only in radiative
$J/\psi$ decays but also in $B \to p \bar p K$, $B^0 \to p \bar \Lambda \pi^-$,
and $B^+ \to p \bar p \pi^+$ decays \cite{Abe:2002ds}, and in $J/\psi \to p
\bar \Lambda K^-$ \cite{Ablikim:2004dj}.  Mechanisms for production of these
enhancements are shown in Fig.\ \ref{fig:bbbar}.  The low-mass $p \bar p$
enhancement seen in $B \to p \bar p K$ appears to be associated with
fragmentation, since antiprotons are seen to follow the $K^+$.  The other
enhancements seen in $B$ decays also appear to favor a fragmentation
mechanism.  The $p \bar \Lambda + \bar p \Lambda$ enhancement seen in $J/\psi$
decays has $M = 2075 \pm 12 \pm 5$ MeV, $\Gamma = 90 \pm 35 \pm 9$ MeV.

\section{LIGHT-QUARK STATES}



The decay $K^+ \to \pi^+ \pi^0 \pi^0$, with a contribution from the
$\pi^+ \pi^+ \pi^-$ intermediate state, measures the $\pi \pi$ scattering
length difference $a_0-a_2$, predicted to be $(0.265\pm0.004)/m_{\pi^+}$
\cite{Cabibbo:2005ez}.  By studying $\pi^+ \pi^- \to \pi^0 \pi^0$
in a $\pi^+ \pi^-$ atom, the DIRAC Collaboration measured $|a_0-a_2| = 
(0.264^{+0.033}_{-0.020})/m_{\pi^+}$ \cite{Adeva:2005pg}.  On the basis of
$2.8 \times 10^7$ $K^+\to \pi^+ \pi^0 \pi^0$ decays and a cleanly observed
cusp in the $M(\pi^0 \pi^0)$ spectrum near $\pi^+ \pi^-$ threshold, the
CERN NA48 Collaboration finds $(a_0-a_2)m_{\pi^+} = 0.281\pm 0.007~({\rm stat})
\pm 0.014~({\rm sys}) \pm 0.014~({\rm theory})$ \cite{Giudici:2005fm}.


As mentioned earlier, a dynamical $I=J=0$ resonance in $\pi \pi \to \pi \pi$ is
predicted using only current algebra, crossing symmetry, and unitarity
\cite{Brown:1971ea}.  A pole with large imaginary part occurs at or below
$m_\rho$.  The effects of this pole are very different in $\pi \pi \to \pi \pi$
[where an Adler zero suppresses the amplitude at low $M(\pi \pi)$] and (e.g.)
$\gamma \gamma \to \pi \pi$, where no such suppression occurs
\cite{Goble:1972rz}.  Evidence for such an effect has been cited in the decays
$D^+ \to \pi^+ \pi^- \pi^+$ \cite{Aitala:2000xu}.  A broad scalar resonance
$\kappa$ near 800 MeV, likely also of dynamical origin, is seen in the
$K^- \pi^+$ system in the decay $D^+ \to K^-\pi^+  \pi^+$, and a
model-independent phase shift analysis shows resonant $J=0$ behavior in
the $K^- \pi^+$ system \cite{Aitala:2002kr}.  Elastic
$K \pi$ scattering does not show similar behavior, casting doubt on the
na\"{\i}ve application of Watson's Theorem to the inelastic process.  The
$\kappa$ is also seen by the BES Collaboration in $J/\psi \to \bar K^{*0}(892)
K^+ \pi^-$ decays \cite{Ablikim:2005ni}.


An effective supersymmetry between scalar $\bar q \bar q$ color triplet states
and spin-1/2 quarks was suggested some time ago \cite{Lichtenberg:1989ix}.
Jaffe and Wilczek \cite{Jaffe:2003sg} have proposed that $\sigma(600)$,
$\kappa(800)$, $f_0(980)$, and $a_0(980)$ could be viewed as a nonet of
diquark--antidiquark pairs.  In that case $J$=0 diquarks would have effective
masses $M([ud]) \simeq 300$ MeV, $M([us]) = M([ds]) \simeq 490$ MeV.  However,
one can also view $\sigma, \kappa, f_0$, and $a_0$ as meson-meson
(``molecular'') effects.  Selem and Wilczek \cite{Wilczek:2004im} have
performed successful fits to baryon Regge trajectories using a quark-diquark
picture.  Interesting features of their treatment are the derivation of a
universal string tension giving a universal slope for trajectories, and a
prediction of the slope for trajectories of baryons with one heavy quark.

If diquarks are really pointlike, the Pauli principle doesn't mind putting
$[ud],~[us],~[ds]$ together in a single hadron, making an ``$H$'' dibaryon
whose mass would be below $2M(\Lambda) = 2.23$ GeV.  The absence of such
a state means that one must be careful about correlations between quarks in
different diquarks.  Shuryak \cite{Shuryak:2005pk} has noted that instanton
interactions allow one to account for such Pauli effects, and lead
to a short-distance repulsive core (0.35 fm) between diquarks.


QCD predicts that in addition to $q \bar q$ states there should be $q \bar q g$
(``hybrid'') states containing a constituent gluon $g$.  One signature of them
would be states with quantum numbers forbidden for $q \bar q$ but allowed for
$q \bar q g$.  For $q \bar q$, $P = (-1)^{L+1},~C = (-1)^{L+S}$, so $CP =
(-1)^{S+1}$.  The forbidden $q \bar q$ states are then those with $J^{PC}=
0^{--}$ and $0^{+-},~1^{-+},~2^{+-},\ldots$.

A consensus in quenched lattice QCD is that the lightest exotic hybrids
have $J^{PC} = 1^{-+}$ and $M(n \bar n g) \simeq 1.9$ GeV, $M(s \bar s g)
\simeq 2.1$ GeV, with errors 0.1--0.2 GeV \cite{McNeile:2002az}.  (Unquenched
QCD opens a Pandora's box of mixing with $qq \bar q \bar q$ and meson pairs.)
Candidates for hybrids include $\pi_1(1400)$ (seen in some $\eta \pi$ final
states, e.g., in $p \bar p$ annihilations) and $\pi_1(1600)$ (seen in $3
\pi$, $\rho \pi$, $\eta' \pi$).

Brookhaven experiment E-852 published evidence for a $1^{-+}$ state called
$\pi_1(1600)$ \cite{Adams:1998ff}.  A recent analysis by a subset of E-852's
participants \cite{Dzierba:2005sr}, however, does not require this particle if
a $\pi_2(1670)$ contribution is assumed.  Such a state would be the Regge
recurrence of the pion, expected in a
picture \cite{Nambu:1974zg} of a rotating QCD string.

\section{CHARMED HADRONS}


CLEO sees $47.2\pm 7.1^{+0.3}_{-0.8}$ events above background of $D^+ \to \mu^+
\nu_\mu$ in $281$ pb$^{-1}$ of $e^+ e^-$ collisions at $E_{\rm cm} = 3.77$ GeV
implying ${\cal B}(D^+ \to \mu^+ \nu_\mu) = (4.40 \pm 0.66 ^{+0.09}_{-0.12})
\times 10^{-4}$ and $f_{D^+} = (222.6 \pm 16.7^{+2.8}_{-3.4})$ MeV
\cite{Artuso:2005ym}.  This is consistent with lattice predictions, including
one \cite{Aubin:2005ar} of $201\pm3\pm17$ MeV.


In the past couple of years the lowest $J^P = 0^+$ and $1^+$ $c \bar s$
states turned out to have masses well below most expectations.  If they had
been as heavy as the already-seen $c \bar s$ states with $L=1$, the
$D_{s1}(2536)$ [$J^P = 1^+$] and $D_{s2}(2573)$ [$J^P = 2^+$]), they would
have been able to decay to $D \bar K$ (the $0^+$ state) and $D^* \bar K$ (the
$1^+$ state).  Instead several groups \cite{Aubert:2003fg} observed a narrow
$D_s(2317)$ decaying to $\pi^0 D_s$ and a narrow $D_s(2460)$ decaying to $\pi^0
D_s^*$.  These decays violate isospin.  Should we have been surprised?

The selection rules in decays of these states show their $J^P$ values
are consistent with $0^+$ and $1^+$.  Low masses are predicted
\cite{Bardeen:2003kt} if these states are viewed as parity-doublets of the
$D_s(0^-)$ and $D^*_s(1^-)$ $c \bar s$ ground states in the framework of
chiral symmetry.  The splitting from the ground states is 350 MeV in each case.
The splitting of the lightest charmed-nonstrange $0^+$ and $1^+$ mesons from
their $0^-$ and $1^-$ partners ($D$ and $D^*$) is a bit larger, as anticipated
in \cite{Bardeen:2003kt}:  $M(0^+)=2308\pm36$ MeV, $\Gamma(0^+)=276 \pm66$ MeV
\cite{Abe:2003zm} or $M(0^+)=2407\pm41$ MeV, $\Gamma(0^+)=240\pm81$ MeV
\cite{Link:2003bd}; $M(1^+) = 2461^{+53}_{-48}$ MeV, $\Gamma(1^+)=
290^{+110}_{-91}$ MeV \cite{Anderson:1999wn} or $M(1^+) = 2427 \pm 36$ MeV,
$\Gamma(1^+)=384^{+130}_{-105}$ MeV \cite{Abe:2003zm}.


The CLEO Collaboration is concerned with finding the best strategy for scanning
$E_{\rm cm}$ in $e^+ e^-$ annihilations to optimize $D_s$ production.  One
can invoke a simple regularity of resonance formation to suggest optimal
energies.  If a meson $M_1$ and a meson $M_2$ have quarks which can mutually
annihilate, they form at least one resonance when $p_{\rm CM} \le 350$ MeV/$c$
\cite{Rosner:1973fq}.  Denote this resonance mass by $M_{\rm max}$. $D \bar D$
states resonate at $\psi''(3770)$, below the predicted value $M_{\rm max}\simeq
3.8$ GeV.  $D^0 \overline{D^{*0}}$ + c.c.\ and $D^+ \overline{D^{*+}}$ + c.c.\
resonances have been seen at 3872 and 3940 MeV.  The $\psi(3S)$ is a
satisfactory candidate for $D^* \overline{D}^*$.  This regularity predicts at
least one $D_s \bar D_s$ resonance below 4.0 GeV and one $D_s \overline{D_s^*}$
+ c.c.\ resonance below 4.08 GeV.  The $Y(4260)$ (see below) is a
candidate for a $D_s^* \overline{D_s^*}$ resonance from this standpoint.
Detailed estimates of higher charmonium production have been performed in
\cite{Barnes:2005pb}.  An old estimate of the cross section shape based on
coupled-channel considerations \cite{Byers:1989zn} would be very useful with
updated $M(D_s)$ and $M(D^*_s)$.


An excited $D_s(2632)$ candidate is seen by the SELEX Collaboration in
$D_s \eta$ (dominant) and $D^0 K^+$ modes \cite{Evdokimov:2004iy}.  In the
first mode, its yield is 40\% that of the $D_s$, which is surprising.  It is
not seen by BaBar, Belle, CLEO, or FOCUS \cite{noDs}.

Doubly-charmed baryon candidates are also seen by SELEX. They are produced by a
(67\% pure) beam $\Sigma^-$ beam.  A $X_{cc}^+(ccd)$ candidate is seen
\cite{Mattson:2002vu} with $M = 3519 \pm 1$ MeV, width resolution-limited,
lifetime $\tau < 33$ fs, decaying to $\Lambda_c^+ K^- \pi^+$.  The signal is
22 events above a background of 6.1.  Surprisingly, the $X_{cc}^+$ yield is
40\% that of $\Lambda_c$.  A weak SELEX signal (5.4/1.6 events) in $D^+ K^- p$
also is seen \cite{Ocherashvili:2004hi}.  There is the claim of an excited
doubly charged state $X_{cc}^{++}(3780)$ decaying to $\Lambda_c K^- \pi^+
\pi^+$, but in contrast to expectations for a strong decay the final state is
not 100\% $X_{cc}^+ \pi^+$.  No other experiments have reported any
doubly-charmed baryon signals.

\begin{figure}
\includegraphics[width=0.98\textwidth]{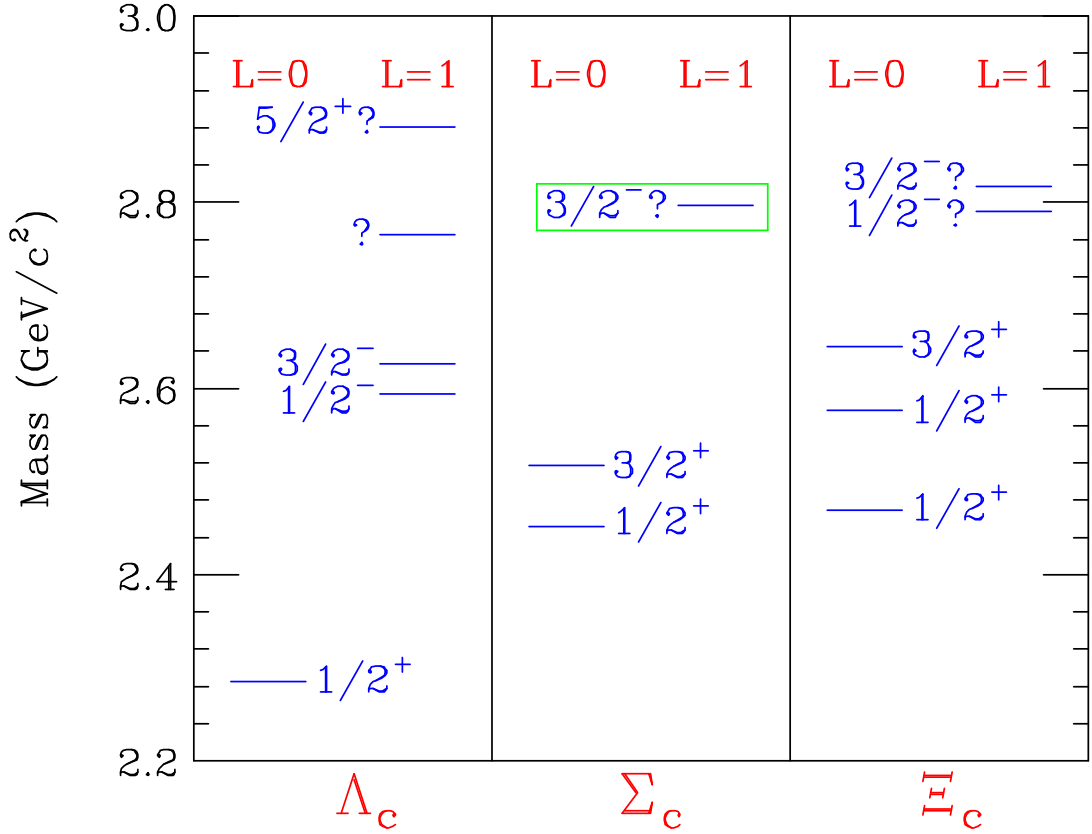}
\caption{Singly-charmed baryons and their excitations.  Box:
state reported in \cite{Mizuk:2004yu}.
\label{fig:lexb}}
\end{figure}

Data on singly-charmed baryons continue to accumulate, following
contributions by CLEO and others.  The reported levels
are summarized in Fig.\ \ref{fig:lexb}.  The Belle Collaboration reports
an excited $\Sigma_c$ candidate decaying to $\Lambda_c \pi^+$, with mass about
510 MeV above $M(\Lambda_c)$ \cite{Mizuk:2004yu}.  It could be an $L=1$
excitation of the spin-1 isotriplet S-wave $uu,ud,dd$ quark pair which, in
the ground state with respect to the $c$ quark, forms the $\Sigma_c$ and
$\Sigma_c^*$.  It could have $J^P = 1/2^-,3/2^-,5/2^-$.  The value shown in
Fig.\ \ref{fig:lexb} is a guess, using the diquark ideas of
\cite{Wilczek:2004im}.  Belle also confirms a $\Lambda_c(2880)$
seen earlier by CLEO.

In Fig.\ \ref{fig:lexb} the first excitations of the
$\Lambda_c$ and $\Xi_c$ are similar, scaling well from the first $\Lambda$
excitations $\Lambda(1405,1/2^-)$ and $\Lambda(1520,3/2^-)$.  They have the
same cost in $\Delta L$, and their $L \cdot S$ splittings scale as $1/m_s$ or
$1/m_c$.  Higher $\Lambda_c$ states may correspond to excitation of a spin-zero
$[ud]$ pair to $S=L=1$, leading to many allowed $J^P$ values up to $5/2^-$.
In $\Sigma_c$ the light-quark pair has $S=1$; adding $L=1$ allows
$J^P \le 5/2^-$.

CLEO has recently remeasured mass differences and widths of the singly-charmed
baryon $\Sigma^*_c(2516, J^P = 3/2^+)$ \cite{Athar:2004ni}.  Mass splittings
between the doubly-charged and neutral states are quite small, in accord with
theoretical expectations.  The prediction of heavy quark symmetry that
$\Gamma(\Sigma_c^{*++})/ \Gamma(\Sigma_c^{++}) = \Gamma(\Sigma_c^{*0})/
\Gamma(\Sigma_c^{0}) = 7.5 \pm 0.1$ is borne out by the data, in which
$\Gamma(\Sigma_c^{*++})/\Gamma(\Sigma_c^{++}) = 6.5 \pm 1.3$,
$\Gamma(\Sigma_c^{*++})/\Gamma(\Sigma_c^{++}) = 7.5 \pm 1.7$.

\section{CHARMONIUM}

The elusive $h_c(1^1P_1)$ state of charmonium has been observed by CLEO
\cite{Rosner:2005ry,Rubin:2005px} via $\psi(2S) \to \pi^0 h_c$ with $h_c \to
\gamma \eta_c$.
Whereas S-wave hyperfine charmonium splittings are $M(J/\psi) - M(\eta_c)
\simeq 115$ MeV for 1S and $M[\psi'] - M(\eta'_c) \simeq $48 MeV for 2S levels,
P-wave splittings should be less than a few MeV since the potential is
proportional to $\delta^3(\vec{r})$ for a Coulomb-like $c \bar c$ interaction.
Lattice QCD \cite{latt} and relativistic potential \cite{Ebert:2002pp}
calculations confirm this expectation of a small P-wave hyperfine
splitting.  One expects $M(h_c) \equiv M(1^1P_1) \simeq
\langle M(^3P_J) \rangle = 3525.36 \pm 0.06$ MeV.

Earlier $h_c$ sightings (see \cite{Rosner:2005ry,Rubin:2005px} for references),
based on
$\bar p p$ production in the direct channel, include a few events at $3525.4
\pm 0.8$ MeV seen in CERN ISR Experiment R704; a state at $3526.2 \pm
0.15 \pm 0.2$ MeV, decaying to $\pi^0 J/\psi$, reported by Fermilab E760 but
not confirmed by Fermilab E835; and a state at $3525.8 \pm 0.2 \pm 0.2$ MeV,
decaying to $\gamma \eta_c$ with $\eta_c \to \gamma \gamma$, reported by
E835 with about a dozen candidate events \cite{Andreotti:2005vu}.

\begin{figure}
\mbox{
\includegraphics[width=0.58\textwidth]{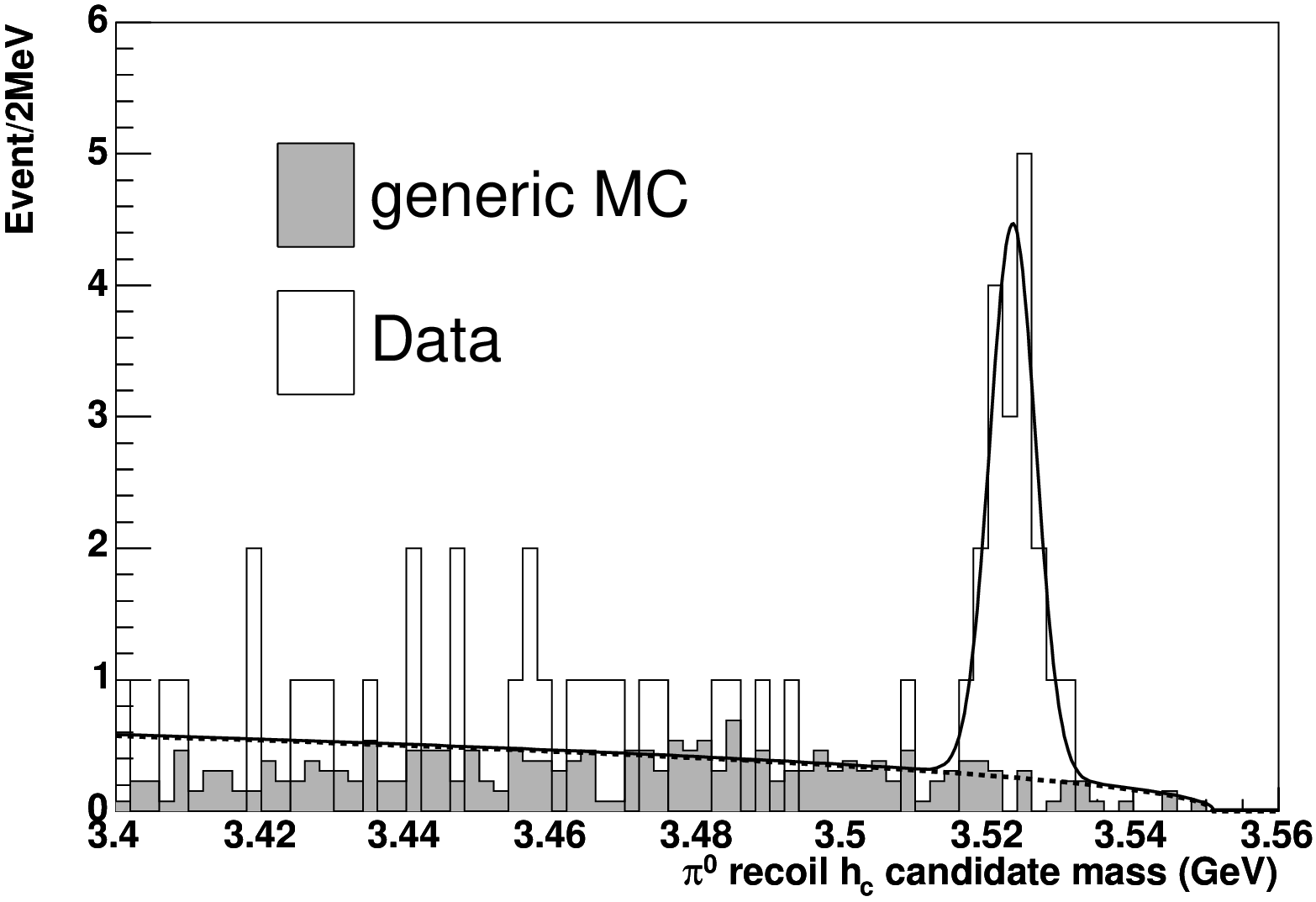}
\includegraphics[width=0.40\textwidth]{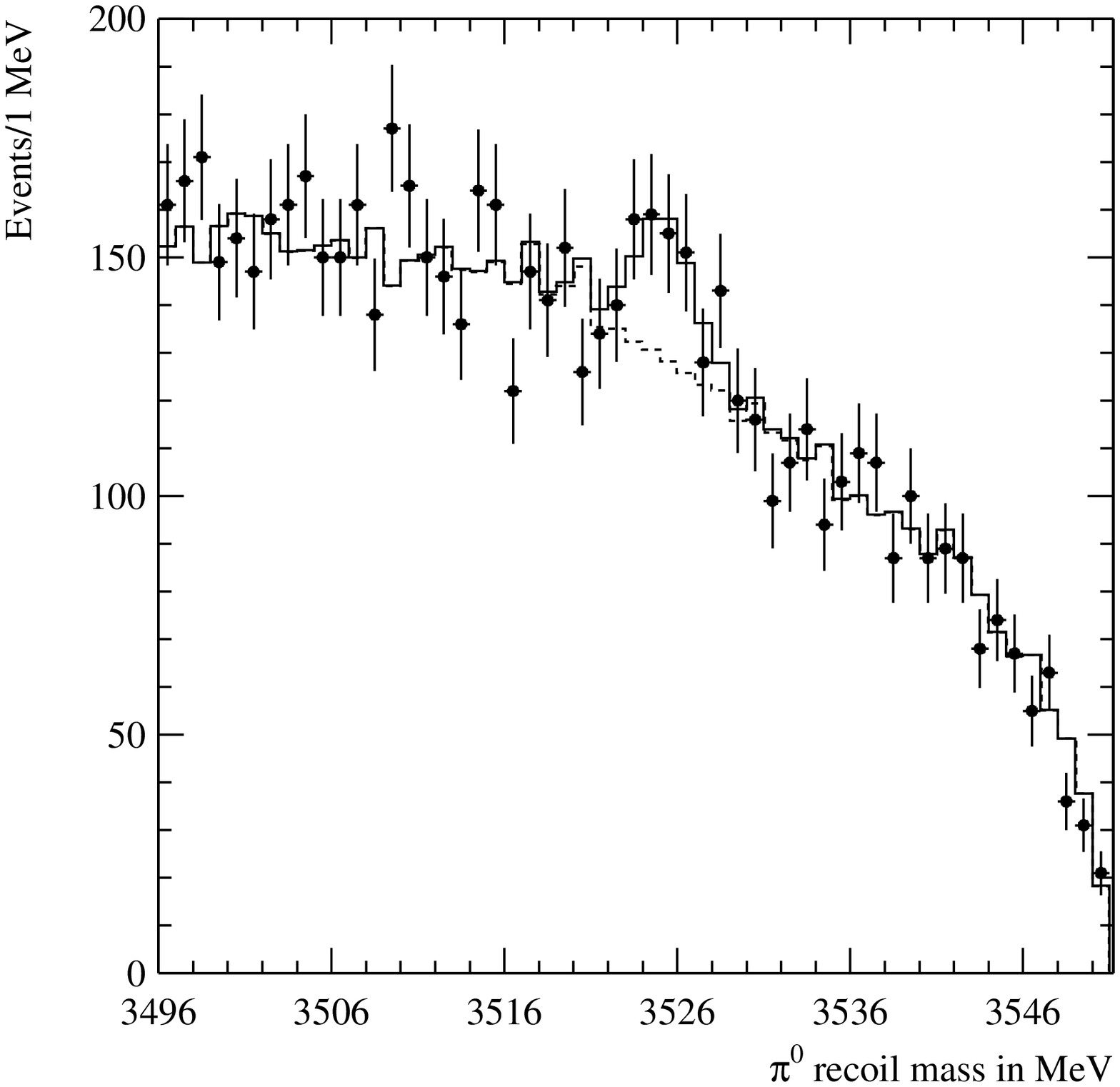}
}
\caption{Left: Exclusive $h_c$ signal from CLEO (3 million $\psi(2S)$
decays).  Data events correspond to open histogram;
Monte Carlo background estimate is denoted by shaded histogram.
The signal shape is a double Gaussian, obtained from signal Monte Carlo.
The background shape is an ARGUS function.
Right: Inclusive $h_c$ signal from CLEO (3 million $\psi(2S)$ decays).
The curve denotes the background function based on generic Monte Carlo plus
signal.  The dashed line shows the contribution of background alone.
Both figures are from Ref.\ \cite{Rubin:2005px}.
\label{fig:hc}}
\end{figure}

In the CLEO data, both inclusive and exclusive analyses see a signal near
$\langle M(^3P_J) \rangle$.  The exclusive analysis reconstructs $\eta_c$ in 7
decay modes, while no $\eta_c$ reconstruction is performed in the inclusive
analysis.
The exclusive signal is shown on the left in Fig.\ \ref{fig:hc}.  A total of 19
candidates were identified, with a signal of $17.5 \pm 4.5$ events above
background.  The mass and product branching ratio for the two transitions 
are $M(h_c) = (3523.6 \pm 0.9 \pm 0.5)$ MeV; ${\cal B}_1(\psi' \to \pi^0 h_c)
{\cal B}_2(h_c \to \gamma \eta_c) = (5.3 \pm 1.5 \pm 1.0) \times 10^{-4}$.
The result of one of two inclusive analyses is shown on the right in
Fig.\ \ref{fig:hc}.  These yield $M(h_c) = (3524.9 \pm 0.7 \pm 0.4)$ MeV,
${\cal B}_1 {\cal B}_2 = (3.5 \pm 1.0 \pm 0.7) \times 10^{-4}$.  Combining
exclusive and inclusive results yields $M(h_c) = (3524.4 \pm 0.6 \pm 0.4)$ MeV,
${\cal B}_1 {\cal B}_2 = (4.0 \pm 0.8 \pm 0.7) \times 10^{-4}$.  The $h_c$ mass
is $(1.0 \pm 0.6 \pm 0.4)$ MeV below $\langle M(^3P_J) \rangle$, barely
consistent with the (nonrelativistic) bound \cite{Stubbe:1991qw} $M(h_c) \ge
\langle M(^3P_J) \rangle$ and indicating little P-wave hyperfine splitting in
charmonium.  The value of ${\cal B}_1 {\cal B}_2$ agrees with theoretical
estimates of $(10^{-3} \cdot 0.4)$.

CLEO has reported a new measurement of $\Gamma(\chi_{c2} \to \gamma \gamma)
=559\pm57\pm45\pm36$ eV based on 14.4 fb$^{-1}$ of $e^+ e^-$ data at $\sqrt{s}=
9.46$--11.30 GeV \cite{chic2}.  The result is compatible with other
measurements when they are corrected for CLEO's new ${\cal B}(\chi_2 \to \gamma
J/\psi)$ and ${\cal B}(J/\psi \to \ell^+ \ell^-)$.  The errors given are
statistical, systematic, and $\Delta {\cal B}(\chi_{c2} \to \gamma J/\psi$).
One can average the CLEO measurement with a corrected Belle result
\cite{Abe:2002va}
to obtain $\Gamma(\chi_{c2} \to \gamma \gamma) = 565 \pm 62$ eV.  Using the
Fermilab E835 value of $\Gamma(\chi_2) = 1.94 \pm 0.13$ MeV
\cite{Andreotti:2005ts}, ${\cal B}(\chi_2 \to \gamma J/\psi) = (19.9 \pm 0.5
\pm1.2)\%$ one finds $\Gamma(\chi_2 \to$ hadrons) = $1.55\pm0.11$ MeV.
This can be compared to $\Gamma(\chi_{c2} \to \gamma \gamma)$, taking
account of QCD radiative corrections \cite{Kwong:1987ak}, to obtain
$\alpha_S(m_c) = 0.290 \pm 0.013$.

CLEO has reported a number of new results based on about 3 million $\psi' =
\psi(2S)$ decays.  The branching fractions for $\psi' \to \gamma X$
have been remeasured, with the results shown in Table \ref{tab:incpsip}
\cite{Athar:2004dn}.  The inclusive $\psi' \to \gamma \chi_{cJ}$ rates are
above the world average \cite{Eidelman:2004wy} values.  The decay $\psi' \to
\gamma \eta_c$ serves as a key calibration for the $\psi' \to \pi^0 h_c \to
\pi^0 \gamma \eta_c$ measurement mentioned above.

\begin{table}
\caption{Branching ratios for $\psi' \to \gamma X$ measured by CLEO
\cite{Athar:2004dn}.
\label{tab:incpsip}}
\begin{tabular}{c c c} \hline
Decay & CLEO ${\cal B}$ (\%) & PDG ${\cal B}$ (\%) \\ \hline
$\psi' \to \gamma \chi_{c2}$ & $9.33 \pm 0.14 \pm 0.61$ & $6.4 \pm 0.6$ \\
$\psi' \to \gamma \chi_{c1}$ & $9.07 \pm 0.11 \pm 0.54$ & $8.4 \pm 0.8$ \\
$\psi' \to \gamma \chi_{c0}$ & $9.22 \pm 0.11 \pm 0.46$ & $8.6 \pm 0.7$ \\
\hline
$\psi' \to \gamma \chi_{cJ}$ & $27.6 \pm 0.3 \pm 2.0$ & $23.4 \pm 1.2$ \\
\hline
$\psi' \to \gamma \eta_c$ & $0.32 \pm 0.04 \pm 0.06$ & $0.28 \pm 0.06$ \\
\hline
\end{tabular}
\end{table}

The decays $\psi' \to J/\psi X$ have also been studied inclusively
\cite{Adam:2005uh}.  The CLEO $\pi^0 \pi^0/\pi^+ \pi^-$ ratio is close to
1/2, as expected from isospin.  The product branching ratios ${\cal B}(\psi'
\to \gamma \chi_{cJ} \to \gamma \gamma J/\psi)$ are found to be larger than
current world averages \cite{Eidelman:2004wy}.  It is found that ${\cal B}
(\psi' \to J/\psi X) = (59.50 \pm 0.15 \pm 1.90)\%$ for $X$ = (all), consistent
with the sum of known modes $(58.9 \pm 0.2 \pm 2.0)\%$.  By subtracting known
modes one can determine ${\cal B}(\psi' \to {\rm~light~hadrons}) = (16.9 \pm
2.6)\%$.  The ratio with respect to a similar $J/\psi$ number
is $2.2 \sigma$ above ${\cal B}(\psi \to \ell^+ \ell^-)/
{\cal B}(J/\psi \to \ell^+ \ell^-) = (12.6 \pm 0.7)\%$ (the ``12\% rule''
expected if decay to light hadrons is due to the same charm-anticharm
annihilation as decay to $\ell^+ \ell^-$).  Strong suppression of some hadronic
$\psi'$ final states, a mystery for more than 20 years, seems
confined to certain species such as $\rho \pi$ and $K^* \bar K$.

The $\psi''(3770)$ is a potential ``charm factory'' for present and future $e^+
e^-$ experiments.  However, the total $\sigma(D \bar D)$ \cite{He:2005bs,%
Ablikim:2004ck} is somewhat less than the average of various direct
measurements of the peak height \cite{Rosner:2004wy,LP452-3}.  To what else
can $\psi''(3770)$ decay, and what might be the reason?  A light-quark
admixture in the $c \bar c$ wave function \cite{Voloshin:2005sd}?
One finds that ${\cal B}(\psi''$ $\pi \pi J/\psi,~\gamma \chi_{cJ}, \ldots)$
sum to at most 1--2\%.  Moreover, both CLEO and BES \cite{LP123}, in searching
for enhanced light-hadron modes, find only that the $\rho \pi$ mode,
suppressed in $\psi(2S)$ decays, also is {\it suppressed} in $\psi''$ decays.
Thus the question of significant non-$D \bar D$ modes of $\psi''$ remains open.

Some branching ratios for $\psi'' \to X J/\psi$ \cite{Adam:2005mr} are
${\cal B}(\psi'' \to \pi^+ \pi^- J/\psi) =(0.214\pm0.025\pm0.022)\%$,
${\cal B}(\psi'' \to \pi^0 \pi^0 J/\psi) =(0.097\pm0.035\pm0.020)\%$,
${\cal B}(\psi'' \to \eta J/\psi) = (0.083\pm0.049\pm0.021)\%$, and
${\cal B}(\psi'' \to \pi^0 J/\psi) < 0.034\%$.
The value of ${\cal B}[\psi''(3770) \to \pi^+ \pi^-
J/\psi]$ found by CLEO is about 2/3 that reported by BES \cite{Bai:2003hv}.
These account for less than 1/2\% of the total $\psi''$ decays.

\begin{table}
\caption{CLEO results on radiative decays $\psi'' \to \gamma \chi_{cJ}$.
Theoretical predictions of Ref.\ \cite{Eichten:2004uh} are (a) without and
(b) with coupled-channel effects; (c) shows predictions of Ref.\
\cite{Rosner:2004wy}.
\label{tab:psipprad}}
\begin{tabular}{|c|c|c|c|c|} \hline
Mode & \multicolumn{3}{c|}{Predicted (keV)} & CLEO (keV) \\ \cline{2-4}
     & (a) & (b) & (c) & preliminary \\ \hline
$\gamma \chi_{c2}$ & 3.2 & 3.9 & 24$\pm$4 & $<40$ (90\% c.l.) \\
$\gamma \chi_{c1}$ & 183 & 59 & $73\pm9$ & $75\pm14\pm13$ \\
$\gamma \chi_{c0}$ & 254 & 225 & 523$\pm$12 & $<1100$ (90\% c.l.) \\ \hline
\end{tabular}
\end{table}

CLEO has recently reported results on $\psi'' \to \gamma \chi_{cJ}$ partial
widths, based on the exclusive process $\psi'' \to \gamma \chi_{c1,2} \to
\gamma \gamma J/\psi \to \gamma \gamma \ell^+ \ell^-$ \cite{Coan:2005}.  The
results are shown in Table \ref{tab:psipprad}.  The exclusive analysis has no
sensitivity to $\chi_{c0}$ since ${\cal B}(\chi_{c0} \to J/\psi)$ is small.
Although the $\psi'' \to \gamma \chi_{c0}$ partial width is expected to be
high, it must be studied in the inclusive channel, which has high background,
or by reconstructing exclusive hadronic $\chi_{c0}$ decays.
Even with the maximum likely $\Gamma(\psi'' \to \gamma \chi_{c0})$, one thus
expects ${\cal B}(\psi'' \to \gamma \chi_{cJ}) < {\cal O}$(2\%).

Several searches for $\psi''(3770) \to ({\rm light~ hadrons})$, including VP,
$K_L K_S$, and multi-body final states, are under way.  The value of
$\sigma(\psi'')$ also is being re-checked.  Two preliminary analyses
\cite{Adams:2005ks,Huang:2005} find no evidence for any light-hadron $\psi''$
mode above expectations from continuum production except $\phi \eta$.
Upper limits on the sum of 26 modes imply ${\cal B}[\psi'' \to$ (light
hadrons)] $\le 1.8\%$.  Cross sections at 3.77 GeV are consistent with
continuum extrapolated from 3.67 GeV, indicating no obvious signature of
non-$D \bar D$ $\psi''$ decays.  Known modes account for at most a few
percent of $\psi''$ decays, to be compared with a possible discrepancy of 1--2
nb in $\sigma(e^+ e^- \to \psi'')$ or ${\cal B} (\psi'') = 10$--20\%.  A
value based on BES data \cite{Rong:2005it}, $\sigma[\psi'' \to ({\rm non-} D
\bar D)] = (0.72\pm0.46\pm 0.62)$ nb does not answer the question.  Could a
more careful treatment of
radiative corrections solve the problem?  A remeasurement of $\sigma(\psi'')$
by CLEO, preferably through an energy scan, is crucial.

\begin{figure}
\includegraphics[width=0.94\textwidth]{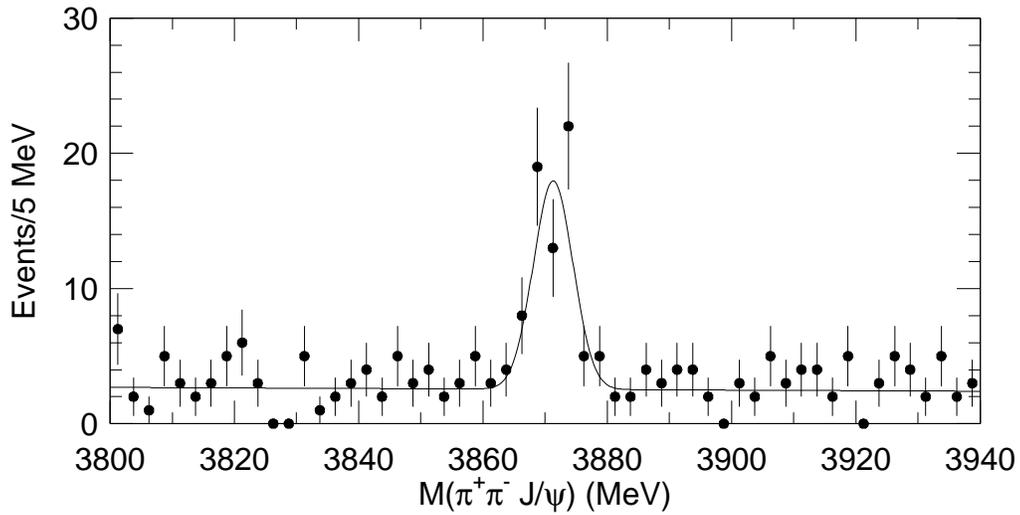}
\caption{Belle distribution in $M(\pi^+ \pi^- J/\psi)$ for the $X(3872)$
region \cite{Abe:2005iy}.
\label{fig:X3872}}
\end{figure}

Many charmonium states above $D \bar D$ threshold have been seen recently.
The $X(3872)$, discovered initially by Belle in $B$ decays \cite{Choi:2003ue}
but confirmed by BaBar \cite{Aubert:2004ns} and in hadronic production
\cite{Acosta:2003zx,Abazov:2004kp}, decays predominantly into $J/\psi \pi^+
\pi^-$.  Evidence for it is shown in Fig.\ \ref{fig:X3872} \cite{Abe:2005iy}.
Since it lies well
above $D \bar D$ threshold but is narrower than experimental resolution (a few
MeV), unnatural $J^P = 0^-,1^+, 2^-$ is favored.  It has many features in
common with an S-wave bound state of $(D^0 \bar D^{*0} + \bar D^0 D^{*0})/
\sqrt{2} \sim c \bar c u \bar u$ with $J^{PC} = 1^{++}$ \cite{Close:2003sg}.
Its simultaneous decay of $X(3872)$ to $\rho J/\psi$ and $\omega J/\psi$ with
roughly equal branching ratios is a consequence of this ``molecular''
assignment.

Analysis of angular distributions \cite{Rosner:2004ac} in $X \to \rho J/\psi,
\omega J/\psi$ favors the $1^{++}$ assignment \cite{Abe:2005iy}.  The detection
of a small $\gamma J/\psi$ mode ($\sim 14\%$ of $J/\psi \pi^+ \pi^-$)
\cite{Abe:2005ix} confirms the assignment of positive $C$ and suggests some
admixture of $c \bar c$ in the wave function.


\begin{figure}
\includegraphics[width=0.98\textwidth]{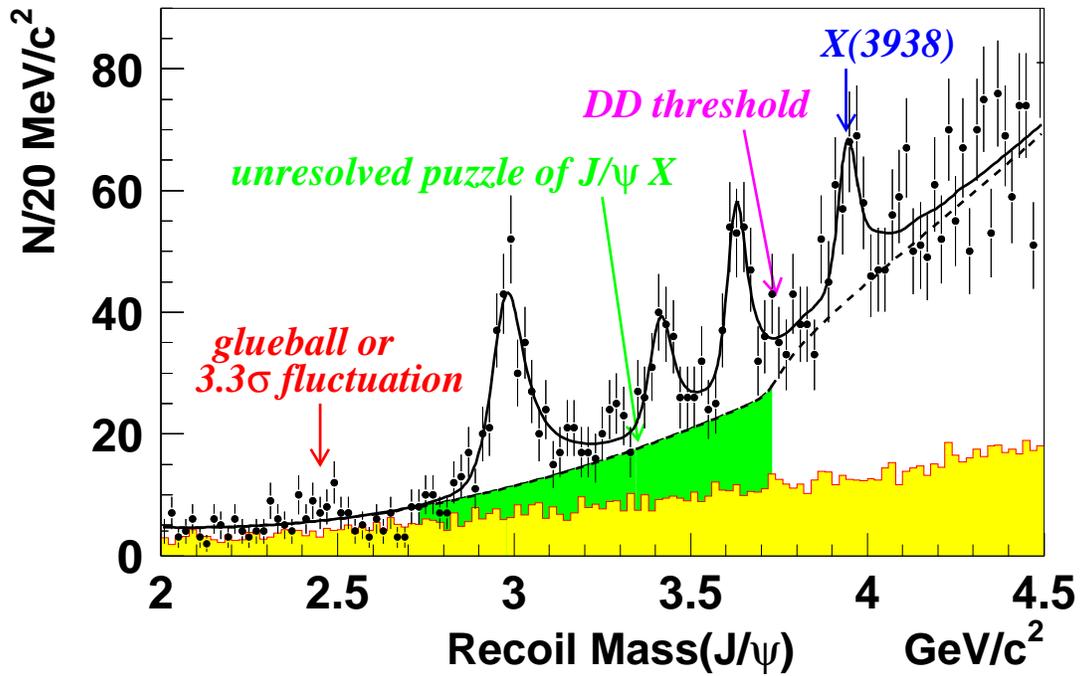}
\caption{Spectrum of masses recoiling against $J/\psi$ in $e^+ e^- \to
J/\psi + X$ \cite{Pakhlov:2004au}.
\label{fig:pakh}}
\end{figure}

The $\omega J/\psi$ final state in $B \to K \omega J/\psi$ shows a peak above
threshold at $M(\omega J/\psi) \simeq 3940$ MeV \cite{Abe:2004zs}.  This could
be a candidate for an excited P-wave charmonium state, perhaps the
$\chi'_{c1,2}(2^3P_{1,2})$.  The corresponding $b \bar b$ states $\chi'_{b1,2}$
have been seen to decay to $\omega \Upsilon(1S)$ \cite{Severini:2003qw}.
A charmonium state distinct from this one but also around 3940 MeV is produced
recoiling against $J/\psi$ in $e^+ e^- \to J/\psi + X$ and is seen to decay to
$D \bar D^*$ + c.c.\ but not $\omega J/\psi$.  Since all lower-mass states
observed in this recoil process have $J=0$ (the $\eta_c(1S), \chi_{c0}$ and
$\eta'_c(2S)$; see Fig.\ \ref{fig:pakh} \cite{Pakhlov:2004au}), it is tempting
to identify this state with $\eta_c(3S)$ (not $\chi'_{c0}$, which would decay
to $D \bar D$).


\begin{figure}
\mbox{\includegraphics[width=0.41\textwidth]{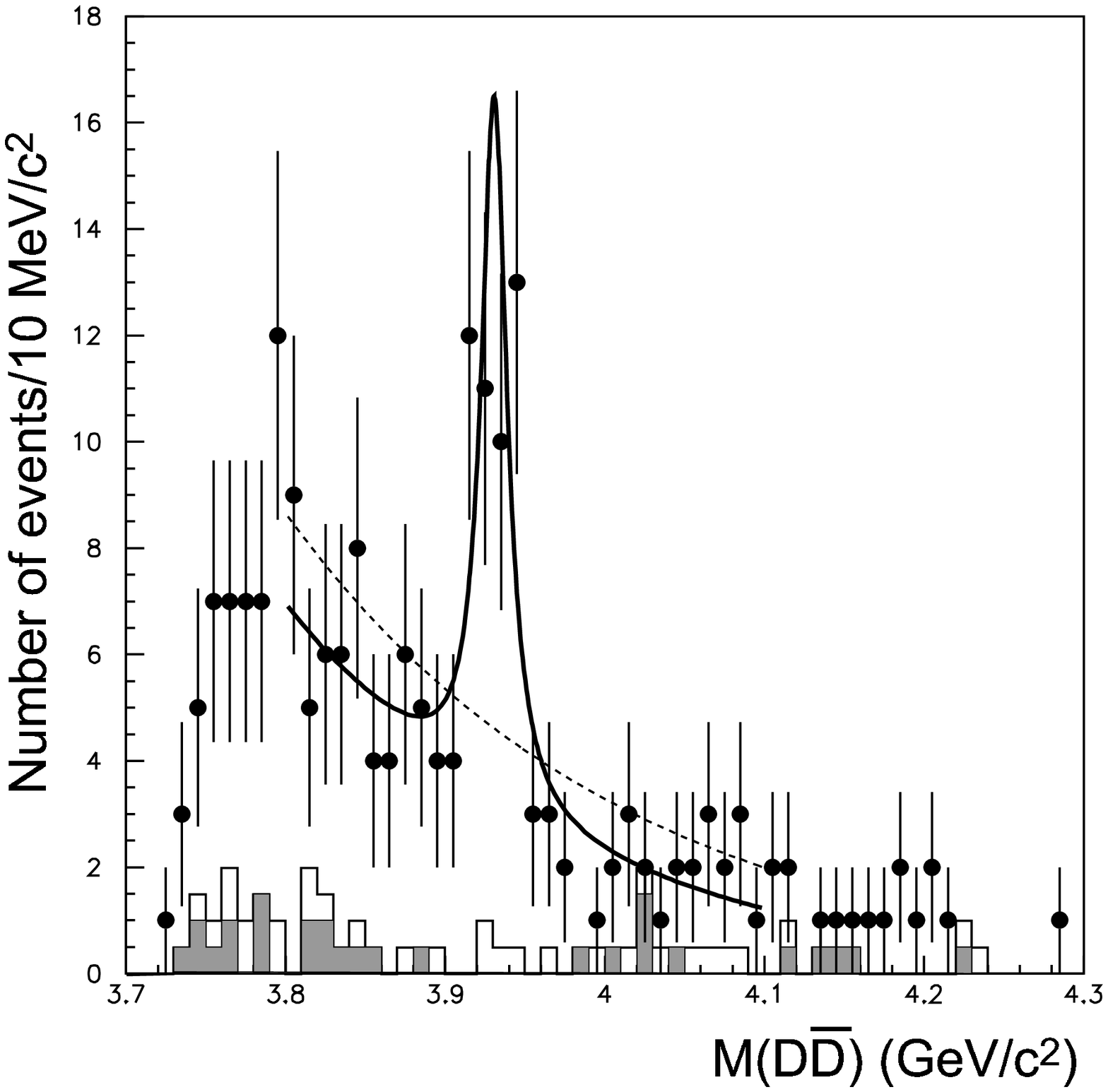}
 \includegraphics[width=0.57\textwidth]{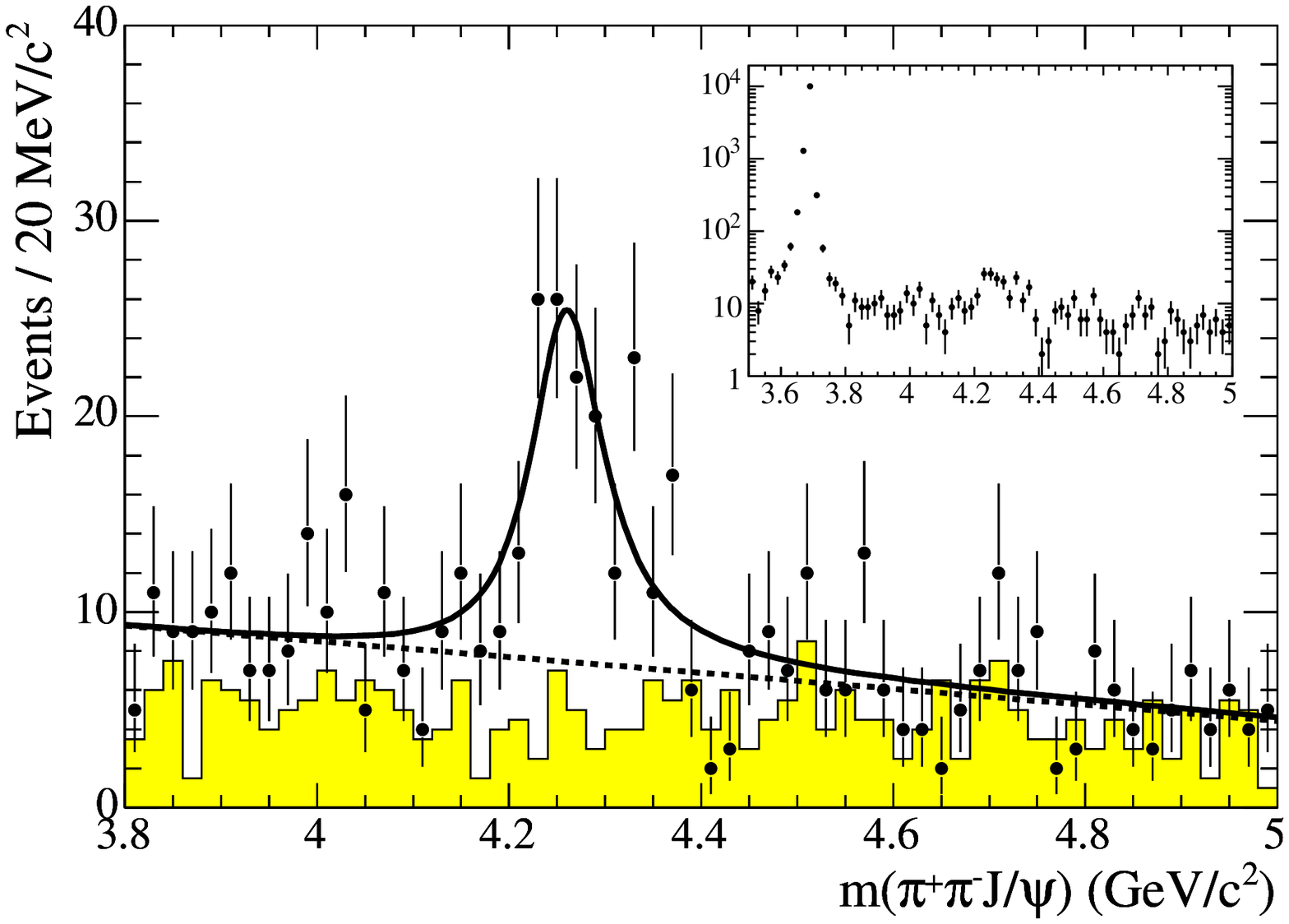}}
\caption{Evidence for excited charmonium states.  Left: $\chi_{c2}(3931)$
(combined $D^0 \bar D^0$ and $D^+ D^-$ spectrum) \cite{Abe:2005bp}.
Right: $Y(4260)$ \cite{Aubert:2005rm}.
\label{fig:spect}}
\end{figure}

Belle has recently reported a candidate for $\chi_{c2}(3931)$ in $\gamma
\gamma$ collisions \cite{Abe:2005bp}, decaying to $D \bar D$.  The spectrum is
shown on the left in Fig.\ \ref{fig:spect}.  The angular distribution of $D
\bar D$ pairs is consistent with $\sin^4 \theta^*$ as expected for a state with
$J=2, \lambda = \pm2$.  It has $M = 3931 \pm 4 \pm 2$ MeV, $\Gamma = 20 \pm 8
\pm 3$ MeV, and $\Gamma_{ee} {\cal B}(D \bar D) = 0.23\pm 0.06 \pm 0.04$, all
of which are reasonable for a $\chi'_{c2}$ state.  Finally, BaBar reports
a state $Y(4260)$ produced in the radiative return reaction $e^+ e^- \to \gamma
\pi^+ \pi^- J/\psi$ and seen in the $\pi^+ \pi^- J/\psi$ spectrum
\cite{Aubert:2005rm}.  Its mass is consistent with being a $4S$ level (see
\cite{Llanes-Estrada:2005vf} for this interpretation) since it lies about 230
MeV above the $3S$ candidate (to be compared with a similar $4S$-$3S$ spacing
in the $\Upsilon$ system).  However, it could also be a hybrid state
\cite{Zhu:2005hp}, as it lies roughly in the expected mass range, a $c s \bar c
\bar s$ state \cite{Maiani:2005pe}, or an effect associated with $D^*_s \bar
D^*_s$ threshold.


\section{BEAUTY HADRONS}

The CDF Collaboration has identified events of the form $B_c \to J/\psi
\pi^\pm$, allowing for the first time a precise determination of the
mass: $M$=6287.0$\pm$4.8$\pm$1.1 MeV \cite{Acosta:2005us}.  This is in
reasonable accord with the latest lattice prediction of
6304$\pm$12$^{+18}_{-0}$ MeV \cite{Allison:2004be}.

We are still waiting for the observation of $B_s$--$\overline{B}_s$ mixing at
the expected level of $\sim 20$ ps$^{-1}$:  The current lower limit
is $> 14.5$ps$^{-1}$ \cite{hfag}.

\section{THE $\Upsilon$ FAMILY (BOTTOMONIUM)}

CLEO data continue to yield new results on $b \bar b$ spectroscopy.
New values of ${\cal B}[\Upsilon(1S,2S,3S) \to \mu^+ \mu^-] = (2.39 \pm 0.02
\pm 0.07, 2.03\pm0.03\pm0.08,2.39\pm0.07\pm0.10)\%$ \cite{Adams:2004xa} imply
lower values of $\Gamma_{\rm tot} (2S,3S)$, which will be important in updating
comparisons with perturbative QCD.  The study of $\Upsilon(2S,3S) \to \gamma X$
decays \cite{Artuso:2004fp} has provided new measurements of E1 transition
rates to $\chi_{bJ}(1P),~\chi'_{bJ}(2P)$ states.  Searches in these data for
the forbidden M1 transitions to spin-singlet states of the form $\Upsilon(n'S)
\to \gamma \eta_b(nS)~(n \ne n')$ have excluded many theoretical models.  The
strongest upper limit, for  $n'=3$, $n=1$, is ${\cal B} \le 4.3 \times 10^{-4}$
(90\% c.l.).  Searches for the lowest $b \bar b$ spin-singlet, the $\eta_b$,
using the sequential processes $\Upsilon(3S) \to \pi^0 h_b(1^1P_1) \to \pi^0
\gamma \eta_b(1S)$ and $\Upsilon(3S) \to \gamma \chi'_{b0} \to \gamma \eta
\eta_b(1S)$ \cite{Voloshin:2004hs} are being conducted.

The direct photon spectrum in $1S,2S,3S$ decays has been measured using CLEO
III data \cite{Besson:2005}.  The ratios $R_\gamma \equiv {\cal B}(g g \gamma)/
{\cal B}(g g g)$ are found to be
$R_\gamma(1S) = (2.50\pm0.01\pm0.19\pm0.13)\%$,
$R_\gamma(2S) = (3.27\pm0.02\pm0.58\pm0.17)\%$,
$R_\gamma(3S) = (2.27\pm0.03\pm0.43\pm0.16)\%$.
$R_\gamma(1S)$ is consistent with an earlier CLEO value
of $(2.54\pm0.18\pm0.14)\%$.

\begin{figure}
\includegraphics[width=0.98\columnwidth]{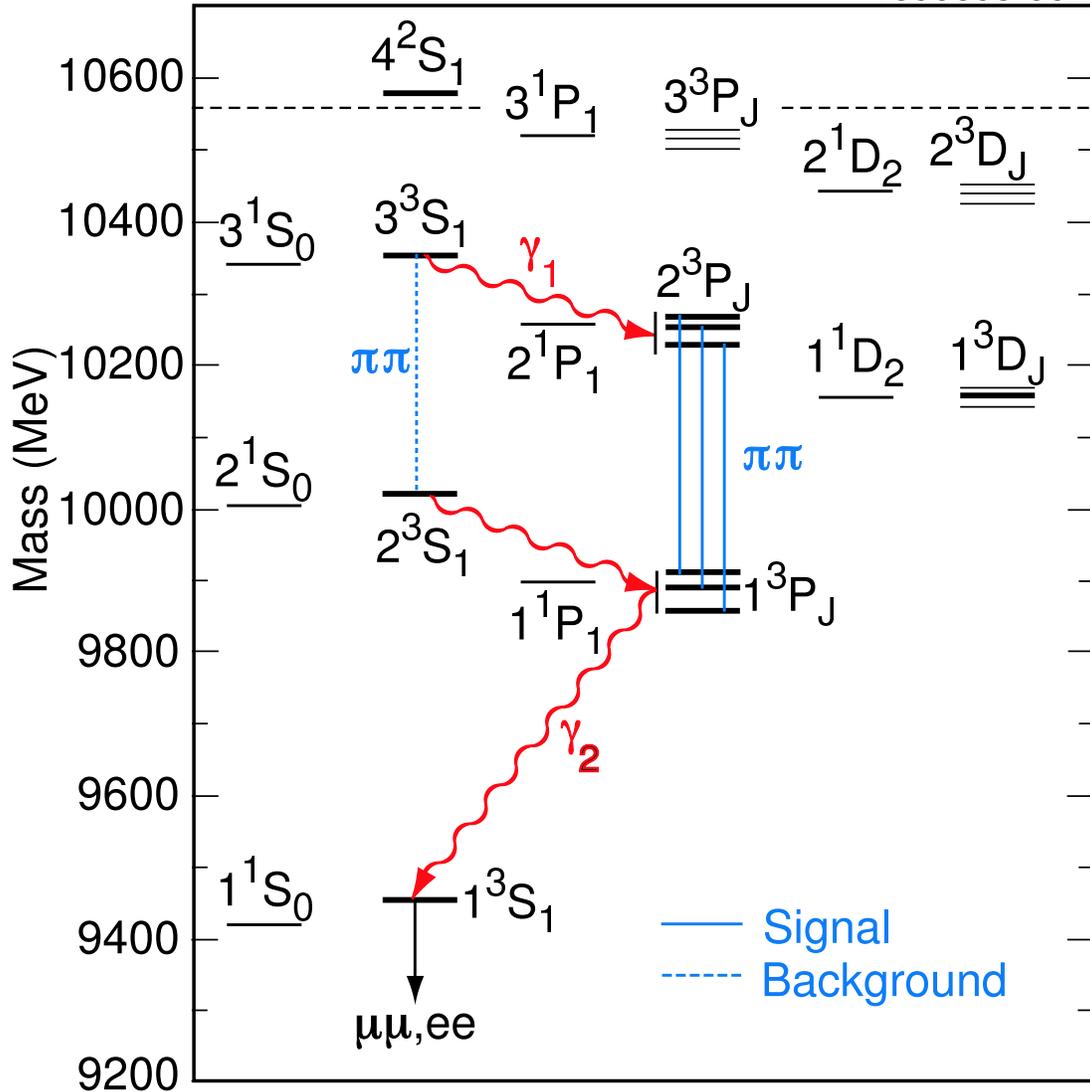}
\caption{Level diagram of $\Upsilon$ states illustrating the transitions
$\chi'_{bJ} \to \pi \pi \chi_{bJ}$.
\label{fig:chipipi}}
\end{figure}

The transitions $\chi'_b \to \chi_b \pi^+ \pi^-$ have been observed for the
first time \cite{Skwarnicki:2005pq}.  One looks for $\Upsilon(3S) \to \gamma
\to \gamma \pi^+ \pi^- \to \gamma \pi^+ \pi^- \gamma \Upsilon(1S)$ in CLEO
data consisting of 5.8 million 3S events.  (See Fig.\ \ref{fig:chipipi}.)
Events with at least one detected soft pion are used.
In the $2 \pi$ sample 7 events are seen above 0.6$\pm$0.2 background  In the
$1 \pi$ sample 17 events are seen above 2.2$\pm$0.6 background.  Assuming
that $\Gamma(\chi'_{b1} \to \pi^+ \pi^- \chi_{b1}) = \Gamma(\chi'_{b2} \to
\pi^+ \pi^- \chi_{b2})$, both are found equal to $(0.80 \pm 0.21
^{+0.23}_{-0.17})$ keV, which is in satisfactory agreement with theoretical
expectations.  Analysis of $\chi'_b \to \pi^0 \pi^0 \chi_b$ is in progress.


\section{A HOMEWORK ASSIGNMENT}

We have made remarkable progress in sorting out level structures and
transition rates for hadrons by appealing to their quark structure and using
the underlying theory based on QCD.  The quarks and leptons themselves present
a much greater challenge.  The underlying source of their masses
and transitions, though these are being mapped out with impressive accuracy,
is unknown.  Will the successes of QCD and the quark model repeat
themselves at another level?  Some suggest that quark and
lepton masses and couplings are as random as the orbits of the planets in the
Solar System.  I hope they are wrong.  Mendeleev spotted gaps in his
periodic table by dealing out elements on cards.  Are we playing with a full
deck of quarks and leptons?  Does our ``periodic table'' have gaps?  Can we
understand it as fundamentally as we understand Mendeleev's?

\section{SUMMARY}

Hadron spectroscopy continues to provide both long-awaited states such as $h_c$
and surprises such as low-lying P-wave $D_s$ mesons and $X(3872)$, $X(3940)$,
$Y(3940)$, $Z(3940)$ (the $\gamma \gamma$ state) and $Y(4260)$.  We continue to
learn how QCD works at strong coupling.  We may have evidence for molecules,
$3S$ and $2P$ candidates, and $4S$ or hybrid charmonium.

QCD may not be the last strongly coupled theory with which we have to deal.
Electroweak symmetry breaking or the very structure of quarks and leptons may
require us to apply similar or related techniques.  These insights largely come
from experiments at the frontier of intensity and detector capabilities rather
than energy, illustrating the importance of a diverse approach to the
fundamental structure of matter.

\begin{theacknowledgments}

I am grateful to J. Appel, T. Browder, F. Close, I. Danko, R. S. Galik,
B. Golob, C. Hearty, H. Mahlke-Kr\"uger, S. Olsen, Y. Sakai, K. Seth, J. Shan,
S. F. Tuan, S. Uehara, B. Yabsley, S. Ye, and J. Yelton for sharing data and
for helpful discussions.  This work was supported in part by the United States
Department of Energy under Grant No.\ DE FG02 90ER40560, and performed in
part at the Aspen Center for Physics. I thank M. Tigner for hospitality of the
Laboratory for Elementary-Particle Physics at Cornell and the John Simon
Guggenheim Foundation for partial support.  

\end{theacknowledgments}

\end{document}